\pdfoutput=1
\documentclass[prd,onecolumn,notitlepage,eqsecnum,nofootinbib,floatfix]{revtex4-1}
\usepackage{amssymb}
\usepackage{amsmath}
\usepackage{bm}
\usepackage{color} 
\usepackage{graphicx}

\begin{document}
\title{Relativistic theory of surficial Love numbers}   
\author{Philippe Landry and Eric Poisson} 
\affiliation{Department of Physics, University of Guelph, Guelph,
  Ontario, N1G 2W1, Canada} 
\date{April 27, 2014} 
\begin{abstract} 
A relativistic theory of surficial Love numbers, which characterize
the surface deformation of a body subjected to tidal forces, was
initiated by Damour and Nagar. We revisit this effort in order to
extend it, clarify some of its aspects, and simplify its computational
implementation. First, we refine the definition of surficial Love
numbers proposed by Damour and Nagar, and formulate it directly in
terms of the deformed curvature of the body's surface, a meaningful
geometrical quantity. Second, we develop a unified theory of surficial
Love numbers that applies equally well to material bodies and black
holes. Third, we derive a compactness-dependent relation between the
surficial and (electric-type) gravitational Love numbers of a
perfect-fluid body, and show that it reduces to the familiar Newtonian
relation when the compactness is small. And fourth, we simplify the
tasks associated with the practical computation of the surficial and
gravitational Love numbers for a material body.    
\end{abstract} 
\maketitle

\section{Introduction and summary} 
\label{sec:intro}

In a famous 1911 treatise \cite{love:11}, the mathematician and
geophysicist A.E.H.\ Love introduced dimensionless quantities that
characterize the deformation of an (otherwise spherical) astronomical
body under the application of tidal forces exerted by remote
bodies. The {\it gravitational Love numbers} $k_\ell$ describe the
deformation of the body's gravitational potential, as measured by the
$\ell$-pole moment of its mass distribution. The {\it surficial Love
  numbers} $h_\ell$ describe the deformation of the body's surface,
also expanded in multipole moments.     

The gravitational and surficial Love numbers provide complementary
information regarding the tidal deformation. Because $k_\ell$ enters
the description of the gravitational field outside the body, it
affects the orbital motion of satellites and other bodies, and the 
gravitational Love numbers are of interest to astronomers. Because
$h_\ell$ is involved in the description of the surface, the surficial
Love numbers are of interest to geophysicists. A measurement of either
set of Love numbers reveals useful information about the body's
internal structure and composition. For a body consisting of a perfect
fluid, the Love numbers are linked by the universal relation 
$h_\ell = 1 + 2 k_\ell$. 

These notions have recently been ported to general relativity. The
tidal deformation of neutron stars has been a topic of active interest
since Flanagan and Hinderer \cite{flanagan-hinderer:08, 
  hinderer:08} pointed out that tidal effects can have measurable
consequences on the gravitational waves emitted by a binary neutron
star in the late stages of its orbital evolution. This implies that the
gravitational waves carry information regarding the internal structure
of each body, from which one can extract useful constraints on the
equation of state of dense nuclear matter. Their initial study was 
followed up with more detailed analyses \cite{hinderer-etal:10,
  baiotti-etal:10, baiotti-etal:11, vines-flanagan-hinderer:11, 
  pannarale-etal:11, lackey-etal:12,damour-nagar-villain:12,
  read-etal:13, vines-flanagan:13, lackey-etal:14, favata:14, yagi-yunes:14}
which conclude that tidal effects might be accessible to measurement
by the current generation of gravitational-wave detectors. 

These observations have motivated the formulation of a proper
relativistic theory of tidal deformations, featuring the precise
definition of relativistic Love numbers. The gravitational Love
numbers $k_\ell$ were promoted to general relativity by Damour and
Nagar \cite{damour-nagar:09} and Binnington and Poisson   
\cite{binnington-poisson:09}, who showed that they actually come in 
two guises: an electric-type variety associated with the
gravito-electric part of the tidal gravitational field, and a 
magnetic-type variety associated with the gravito-magnetic
interaction. The gravitational Love numbers were computed for 
relativistic polytropes \cite{damour-nagar:09, binnington-poisson:09},
and for realistic neutron-star models constructed from tabulated
equations of state \cite{hinderer-etal:10, postnikov-etal:10,
  pannarale-etal:11, damour-nagar-villain:12}. The gravitational Love 
numbers of a black hole were shown to be precisely zero 
\cite{binnington-poisson:09}. 

The gravitational Love numbers of neutron stars have been implicated
in a set of relations --- the $I$-Love-$Q$ relations 
\cite{yagi-yunes:13a, yagi-yunes:13b, doneva-etal:13, maselli-etal:13,
  yagi:14, haskell-etal:14}  --- involving the moment of
inertia $I$, the Love number $k^{\rm el}_2$, and the quadrupole moment
$Q$ of a neutron star. While each quantity depends on the star's
internal structure and composition, particular relations between these  
quantities display a remarkable independence relative to the details
of internal structure.     

The surficial Love numbers $h_\ell$ were also promoted to general
relativity by Damour and Nagar \cite{damour-nagar:09}. In this paper
we revisit this effort in order to extend it, clarify some of its
aspects, and simplify its computational implementation.  

Our first goal is to provide a refined definition of the
surficial Love numbers as coordinate-invariant quantities in general
relativity. In the original definition proposed by Damour and Nagar,
the Love numbers are related to the coordinate deformation $\delta R$
of the body's surface under the application of a tidal force. This 
definition is imported from the Newtonian theory, and it is made
geometrically meaningful by embedding the two-dimensional surface in a
three-dimensional, Euclidean space, so that $\delta R$ can be related
to the curvature of this surface. Here we proceed differently. Instead
of introducing a fictitious Euclidean space in the relativistic
theory, we begin by formulating an alternative definition of the
Newtonian Love numbers in terms of a curvature perturbation 
$\delta {\cal R}$ instead of a surface displacement $\delta R$. The
relation between $\delta {\cal R}$ and $\delta R$ is still obtained by
embedding the body's surface in a Euclidean space, but this is now
done in the Newtonian theory, for which the Euclidean manifold is a
key foundational element. The new definition for the Newtonian love
numbers is then promoted to general relativity by preserving the
Newtonian relation between $\delta {\cal R}$ and the tidal field. Our
relativistic definition is equivalent to the one proposed by Damour
and Nagar, but we feel that it is more compelling from a geometrical
point of view. In essence, our approach is to geometrize the Newtonian 
definition by relying on the pre-existing Euclidean manifold, and to
promote this definition to general relativity; their approach is to
promote the coordinate definition first, and then to geometrize it 
with the help of a fictitious Euclidean space introduced as an
additional structure.    

Our second goal is to formulate a unified theory of relativistic
surficial Love numbers that applies equally well to material bodies
and (nonrotating) black holes. In their original work, Damour and
Nagar noticed that the surficial Love numbers of a black hole, which
were previously calculated by Damour and Lecian
\cite{damour-lecian:09}, can be recovered in the limit when the
compactness $M/R$ of a material body approaches $1/2$; $M$ is the
body's mass, and $R$ is its areal radius. Because the limit cannot be
physically attained by material bodies in hydrostatic equilibrium, the
meaning of this coincidence was left unclear in the original work. We
aim to provide clarification by showing that our geometrical
definition for the surficial Love numbers applies to both material
bodies and black holes, and that the equation relating $h_\ell$ to the
metric perturbations takes the same form in both cases. 

\begin{figure} 
\includegraphics[width=0.8\linewidth]{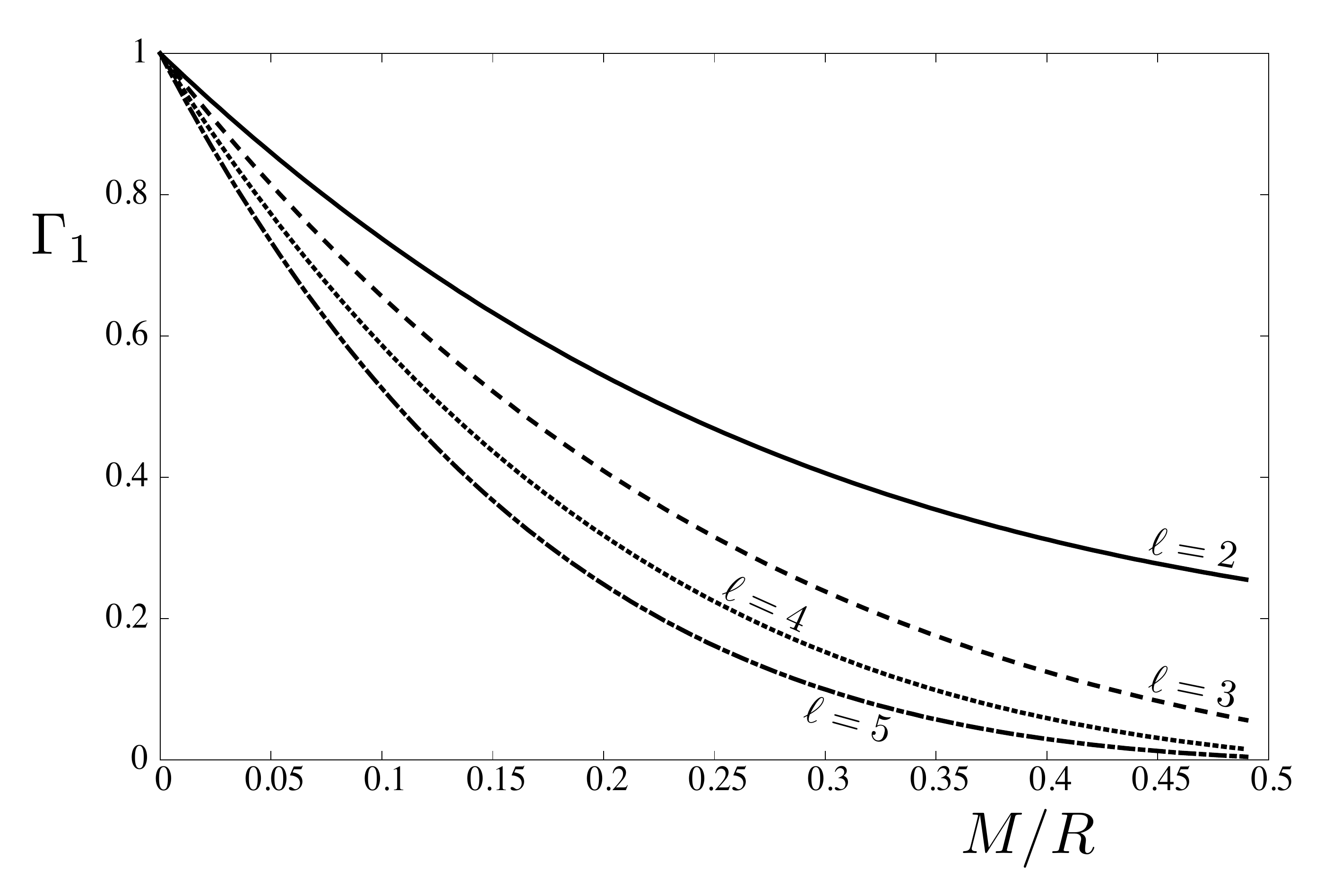}
\caption{Plot of the coefficient $\Gamma_1$ as a function of
  compactness $M/R$, for selected values of $\ell$.} 
\label{fig:f1} 
\end{figure} 

\begin{figure} 
\includegraphics[width=0.8\linewidth]{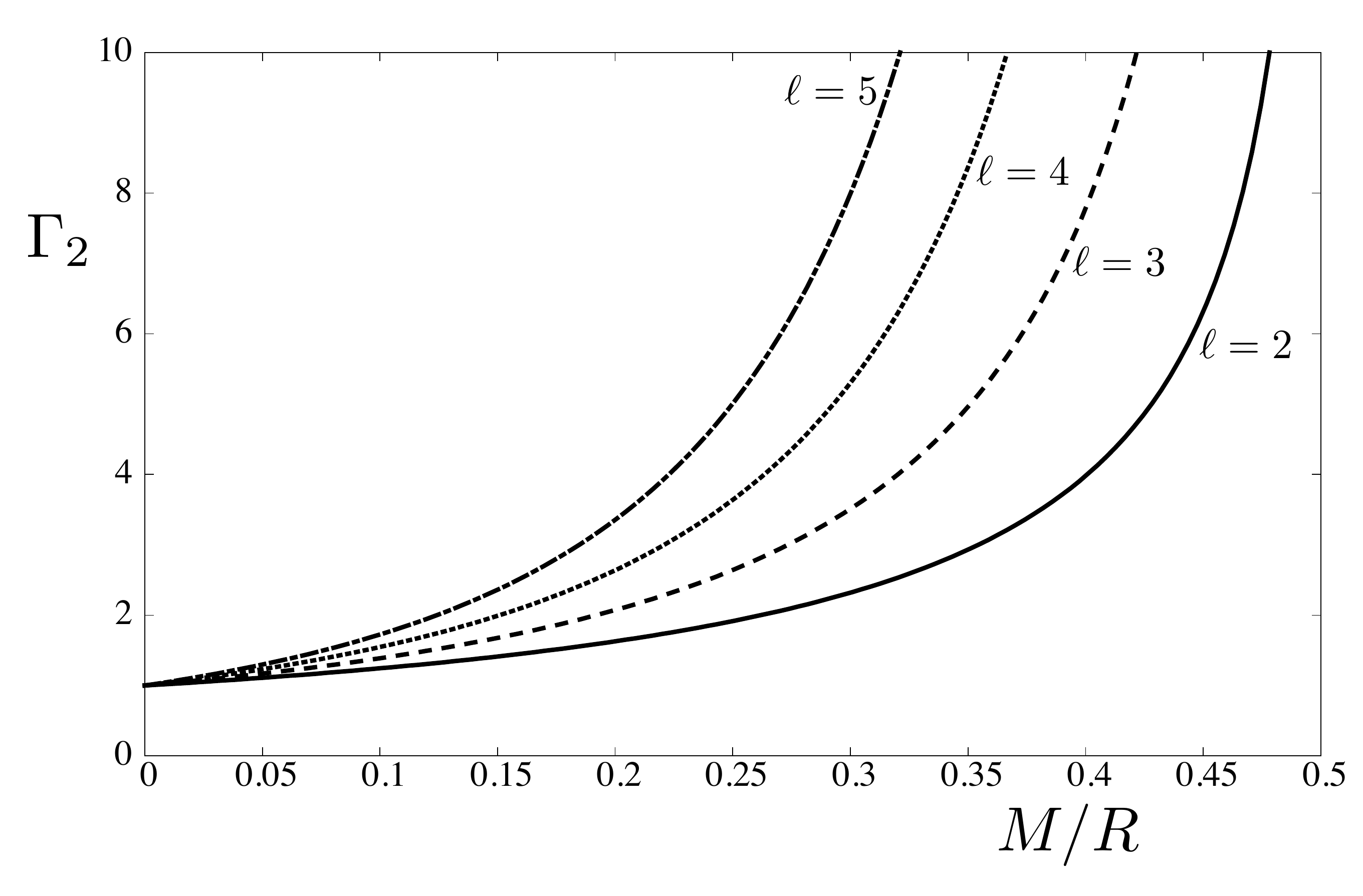}
\caption{Plot of the coefficient $\Gamma_2$ as a function of
  compactness $M/R$, for selected values of $\ell$.} 
\label{fig:f2} 
\end{figure} 

Our third goal is to derive a simple universal relation between the
surficial and (electric-type) gravitational Love numbers of a
perfect-fluid body; such a relation was not noticed in the original
work by Damour and Nagar. The relation is 
\begin{equation} 
h_\ell = \Gamma_1 + 2  \Gamma_2\, k^{\rm el}_{\ell}, 
\label{h_vs_k1} 
\end{equation} 
and it is universal in the sense that the coefficients $\Gamma_1$ and
$\Gamma_2$ are functions of the compactness $M/R$ only; they are
independent of all other details of internal structure. They are given
by  
\begin{subequations} 
\label{Gamma1} 
\begin{align} 
\Gamma_1 &= \frac{\ell+1}{\ell-1} (1-M/R) F(-\ell,-\ell;-2\ell;2M/R) 
- \frac{2}{\ell-1} F(-\ell,-\ell-1;-2\ell;2M/R), \\ 
\Gamma_2 &= \frac{\ell}{\ell+2} (1-M/R) F(\ell+1,\ell+1;2\ell+2;2M/R) 
+ \frac{2}{\ell+2} F(\ell+1,\ell;2\ell+2;2M/R), 
\end{align} 
\end{subequations} 
where $F(a,b;c;z)$ is the hypergeometric function.  A relationship of
the form of Eq.~(\ref{h_vs_k1}) was first written down by Yagi
\cite{yagi:14}, but the expression provided here is substantially
simpler. The figures reveal that $\Gamma_1$ is a decreasing function
of $M/R$, while $\Gamma_2$ is an increasing function. When 
$M/R \ll 1$ the functions can be approximated by  
\begin{subequations} 
\label{Gamma_series1} 
\begin{align} 
\Gamma_1 &= 1 - (\ell+1) (M/R) 
+ \frac{ \ell(\ell+1)(\ell^2-2\ell+2) }{ (\ell-1)(2\ell-1) } (M/R)^2 
+ \cdots, \\ 
\Gamma_2 &= 1 + \ell (M/R) 
+ \frac{ \ell(\ell+1)(\ell^2+4\ell+5) }{ (\ell+2)(2\ell+3) } (M/R)^2 
+ \cdots. 
\end{align} 
\end{subequations} 
In the limit $M/R \to 0$ we recover the Newtonian relation, 
$h_\ell = 1 + 2 k^{\rm el}_\ell$.  

Because it is issued from the unified theory, the relation of
Eq.~(\ref{h_vs_k1}) can be applied directly to black holes, for which
$M/R = 1/2$ and $k^{\rm el}_\ell = 0$. Evaluation of the
hypergeometric functions reveals that the surficial Love numbers of a
nonrotating black hole are given by  
\begin{equation} 
h_\ell = \frac{\ell+1}{2(\ell-1)} \frac{\ell!^2}{(2\ell)!}.
\label{h_BH1} 
\end{equation}
This agrees with the result of Damour and Lecian
\cite{damour-lecian:09}.    
 
Our fourth goal is to simplify the tasks associated with the practical
computation of the (electric-type and magnetic-type) gravitational
Love numbers for a material body governed by an arbitrary equation of
state, in the hope to facilitate future investigations. These
simplified procedures can also benefit the computation of surficial
Love numbers, which are obtained from Eq.~(\ref{h_vs_k1}).   

The paper is organized as follows. We begin in Sec.~\ref{sec:Newton}
with a review of Love numbers in the Newtonian theory of tidal
deformations. To prepare the way for a relativistic definition of
surficial Love numbers, we translate the traditional Newtonian
definition, formulated in terms of a displacement of the surface, to
an equivalent definition involving the perturbation of its intrinsic
(Gaussian) curvature. This Newtonian definition is promoted to general
relativity in Sec.~\ref{sec:relativity}. In the following sections we
endeavor to calculate the surficial Love numbers for
perfect-fluid bodies and black holes. In Sec.~\ref{sec:hydrostatic} we
describe the perturbed hydrostatic equilibrium of a tidally deformed
body, and establish that the surficial Love numbers are
gauge-invariant quantities. In Sec.~\ref{sec:blackhole} we show that
the main results obtained in Sec.~\ref{sec:hydrostatic} apply just as
well to a tidally deformed black hole, in spite of the fact that the
surface of a material body and the event horizon of a black hole are
physically very different. This observation implies that the surficial
Love numbers of all compact bodies can be defined in a unified way. In 
Sec.~\ref{sec:h_vs_k} we derive the relation of Eq.~(\ref{h_vs_k1}),
and in Sec.~\ref{sec:black} we establish Eq.~(\ref{h_BH1}). In
Sec.~\ref{sec:recipe} we review and refine the method
devised by Damour and Nagar \cite{damour-nagar:09} to calculate 
the (electric-type and magnetic-type) gravitational Love numbers, a
necessary input into the computation of the surficial Love numbers.  

In this paper we complete the task, initiated by Damour and Nagar, of
providing Love numbers with a proper relativistic formulation. The
physical significance of gravitational Love numbers has been clearly
demonstrated in the work reviewed above, and Tsang {\it et al.}\ have
begun an exploration of the importance of surficial Love numbers in
astrophysical processes involving neutron stars
\cite{tsang-etal:12}. Our interest in carrying out this work, however, 
stems mostly from the observation that the Newtonian theory of tidal
deformations and dynamics has achieved a great degree of perfection in
the last three hundred years, in addition to delivering a host of
highly accurate results. We aspire to the same degree of completeness
and perfection for the relativistic theory.      

\section{Love numbers in Newtonian theory} 
\label{sec:Newton} 

(Our treatment below follows Secs.~2.4 and 2.5 of
Ref.~\cite{poisson-will:14}, except for a change of normalization for
the tidal moments.)  

We consider, within the context of Newtonian gravity, a spherical body
of mass $M$ and radius $R$ that is slightly deformed by tidal forces
exerted by remote bodies. Working in the moving (and noninertial)
reference frame of the body, the external potential is conveniently
expressed as a Taylor expansion about the body's center-of-mass. We
have   
\begin{equation} 
U_{\rm ext}(\bm{x}) = -\sum_{\ell=2}^\infty \frac{1}{(\ell-1)\ell} 
r^\ell {\cal E}_L \Omega^L, 
\label{Uext} 
\end{equation} 
where $\bm{x}$ is the displacement from the center-of-mass, 
$r := |\bm{x}|$, $\bm{\Omega} := \bm{x}/r 
= [\sin\theta\cos\phi, \sin\theta\sin\phi, \cos\theta]$, 
$L := a_1 a_2 \cdots a_\ell$ is a multi-index containing a number
$\ell$ of individual indices, and $\Omega^L := \Omega^{a_1}
\Omega^{a_2} \cdots \Omega^{a_\ell}$ is a string of $\ell$ angular
vectors; summation over a repeated multi-index implies summation over
all individual indices. The tensors 
\begin{equation} 
{\cal E}_L := -\frac{1}{(\ell-2)!} \partial_L U_{\rm ext}(\bm{0}),  
\label{E_Ndef} 
\end{equation} 
where $\partial_L := \partial_{a_1} \partial_{a_2}
\cdots \partial_{a_\ell}$, are {\it tidal moments} associated with
the external field. The tensors ${\cal E}_L$ are symmetric
in all their indices, and they are also completely tracefree (a
contraction of any pair of indices vanishes) by virtue of the fact
that the external potential satisfies Laplace's equation 
$\nabla^2 U_{\rm ext} = 0$ in the vicinity of the reference body. A
factor of $(\ell-2)!$ was inserted in Eq.~(\ref{E_Ndef}) to make
contact with the relativistic description of the tidal environment, to
be introduced below.  

The decomposition of the external potential in
symmetric-tracefree (STF) tensors is mathematically equivalent to a
decomposition in spherical harmonics (see Sec.~1.5 of
Ref.~\cite{poisson-will:14}). It can indeed be shown
that the STF tensor $\Omega^{\langle L \rangle}$ (with the angular
brackets indicating the operation of trace removal on all indices) is
in a one-to-one correspondence with the spherical harmonics of degree 
$\ell$. As a consequence we may write 
\begin{equation} 
{\cal E}_L \Omega^L =  \sum_{\ell = -m}^\ell 
{\cal E}^{(\ell)}_m Y_{\ell m}(\theta,\phi), 
\label{STF_Ylm} 
\end{equation} 
with the $2\ell + 1$ coefficients ${\cal E}^{(\ell)}_m$ providing a
packaging of the  $2\ell+1$ independent components of ${\cal E}_L$. 

We consider situations for which the time dependence contained in the 
tidal moments is sufficiently slow that the tidal forces never take the
body out of equilibrium. In such situations, the external time scale
(comparable to $\sqrt{a^3/(GM)}$, in which $a$ is the inter-body
distance) is much longer than the internal time scale (comparable to
$\sqrt{R^3/(GM)}$) attached to physical processes taking place inside
the body. In this regime the relaxation to a new equilibrium state
following a slow change in the tidal environment (caused by the
orbital motion of the remote bodies) can be idealized as taking place 
instantaneously. The time dependence of ${\cal E}_L$ is therefore
irrelevant to the internal dynamics, and it can be ignored for the
purposes of calculating the body's deformation; although the tidal
moments carry a parametric dependence upon time, they can still be  
treated as constants in our developments. This defines the regime of 
{\it static tides}. 

The tidal forces produced by the remote bodies create a slight
deformation of the reference body. One way of measuring the
deformation is through the multipole moments of its mass distribution,
defined by $I^L := \int \rho r^\ell \Omega^{\langle L \rangle}\, d^3x$. 
These would all vanish in spherical symmetry, but in the presence of
tidal forces we have instead 
\begin{equation} 
I_L = -\frac{ 2(\ell-2)!}{(2\ell-1)!! } k_\ell R^{2\ell+1} {\cal E}_L,  
\label{k_Ndef} 
\end{equation} 
where $k_\ell$, the {\it gravitational Love numbers}, are
dimensionless and scale-free measures of the deformation; these
quantities depend on the details of internal structure, and are
computed on the basis of a detailed description of the perturbed
interior. The total gravitational potential can then be written as 
$U = M/r + \delta U$, where 
\begin{equation} 
\delta U = -\sum_{\ell=2}^\infty \frac{1}{(\ell-1)\ell} 
\Bigl[ 1 + 2 k_\ell (R/r)^{2\ell+1} \Bigr] 
r^\ell {\cal E}_L \Omega^L, 
\label{U_tot} 
\end{equation} 
with the first set of terms (growing as $r^\ell$) representing the
external potential, and the second set (decaying as $r^{-(\ell+1)}$)
representing the body's response. 

Another way of measuring the tidal deformation is to examine the
body's surface. In the absence of a perturbation the surface is
spherical and described by the equation $r = R$. In the presence
of a deformation we have instead $r = R + \delta R$, with 
\begin{equation} 
\delta R = -\sum_{\ell=2}^\infty \frac{h_\ell}{(\ell-1)\ell} 
\frac{R^{\ell+2}}{M} {\cal E}_L \Omega^L, 
\label{h_Ndef1} 
\end{equation} 
where $h_\ell$, the {\it surficial Love numbers}, are other
dimensionless and scale-free measures of the deformation. These also 
depend on the details of internal structure. When the body consists 
of a perfect fluid, the body's surface is necessarily an
equipotential, and evaluating $U$ at $r= R + \delta R$
implies that the Love numbers are related by 
\begin{equation} 
h_\ell = 1 + 2 k_\ell. 
\label{hvsk_N} 
\end{equation} 
The relation is violated when the body's interior is not well modeled
as a fluid; this happens, for example, when there is a solid mantle
surrounding a fluid core.      

Because Eq.~(\ref{h_Ndef1}) is formulated in terms of a coordinate
displacement, it is not immediately amenable to a relativistic
generalization, which should provide $h_\ell$ with a
coordinate-invariant meaning. It is easy, however, to turn
Eq.~(\ref{h_Ndef1}) into a geometrical statement that can be promoted
to general relativity. The key is to describe the deformation not in
terms of the surface displacement, but equivalently in terms of the
perturbation in the surface's intrinsic curvature. 

To effect this translation we examine the intrinsic geometry of a
surface $r = R + \delta R$ in flat space. Working with the angular
coordinates $\theta^A = (\theta,\phi)$, we find that to first order in
the displacement, the metric on the deformed surface is given
by $ds^2 = R^2 (1 + 2F) d\Omega^2$, in which $F := \delta R/R$ and
$d\Omega^2 = d\theta^2 + \sin^2\theta\, d\phi^2$. This is a special
case of 
\begin{equation} 
ds^2 = R^2 ( \Omega_{AB} + \delta \Omega_{AB} ) d\theta^A d\theta^B, 
\label{deformed_sphere1} 
\end{equation} 
in which $\Omega_{AB} := \mbox{diag}[1,\sin^2\theta]$, the metric on a
round two-sphere of unit radius, is given a deformation 
$\delta \Omega_{AB}$. It is easy to show that to first order in the
deformation, the Ricci scalar of the two-dimensional surface is given
by 
\begin{equation}  
{\cal R} = \frac{1}{R^2} ( 2 + \delta {\cal R}), 
\label{deformed_sphere2} 
\end{equation} 
with 
\begin{equation} 
\delta {\cal R} = \bigl[ D^A D^B 
- (D^2 + 1) \Omega^{AB} \bigr] \delta \Omega_{AB}, 
\label{deformed_sphere3} 
\end{equation} 
where $D_A$ is the covariant derivative compatible with $\Omega_{AB}$
(so that $D_A \Omega_{BC} = 0$), $D^A := \Omega^{AB} D_B$ with
$\Omega^{AB}$ the matrix inverse to $\Omega_{AB}$, and 
$D^2 := \Omega^{AB} D_A D_B$ is the Laplacian operator on the unit
two-sphere. Notice that in Eq.~(\ref{deformed_sphere3}), all tensorial
operations are performed on a unit two-sphere with metric
$\Omega_{AB}$ instead of a two-sphere of radius $R$.   

In our specific case we have $\delta \Omega_{AB} = 2F \Omega_{AB}$,
and the curvature perturbation reduces to $\delta {\cal R} = -2(D^2+2)
F$. With $F$ given by Eq.~(\ref{h_Ndef1}), substitution of the
spherical-harmonic decomposition of Eq.~(\ref{STF_Ylm}) produces 
\begin{equation} 
\delta {\cal R} = -2\sum_{\ell=2}^\infty \frac{\ell+2}{\ell} h_\ell  
\frac{R^{\ell+1}}{M} {\cal E}_L \Omega^L, 
\label{h_Ndef2} 
\end{equation} 
an alternative definition for the surficial Love numbers. To arrive at
this result we invoked the eigenvalue equation for spherical
harmonics, $D^2 Y_{\ell m} = -\ell(\ell+1) Y_{\ell m}$. 

\section{Surficial Love numbers in general relativity} 
\label{sec:relativity} 

With a suitable reinterpretation of $\delta {\cal R}$ on the left-hand
side, and ${\cal E}_L$ on the right-hand side, Eq.~(\ref{h_Ndef2}) can
be promoted to a precise definition of surficial Love numbers in
general relativity. As we shall show, this definition possesses the
essential property that the Love numbers are coordinate-invariant
quantities.   

Zhang \cite{zhang:86} provided a characterization of a tidal 
environment in general relativity with two sets of tidal
moments ${\cal E}_L$ and ${\cal B}_L$. These are defined in terms of
the asymptotic behavior of the spacetime's Weyl tensor far away from 
the reference body (where $M \ll r \ll a$). Zhang's expression for
$g_{tt}$, specialized to a weak-field situation, is compatible with
Eq.~(\ref{Uext}) with the normalization of the tidal moments specified
in Eq.~(\ref{E_Ndef}). Zhang's construction, however, allows us to
define the tidal moments ${\cal E}_L$ in full general relativity, and
therefore to provide a meaningful interpretation for the right-hand
side of Eq.~(\ref{h_Ndef2}). The second set of tidal moments, 
${\cal B}_L$, does not enter in the definition of surficial Love
numbers in general relativity.  

The boundary of a material body defines a three-dimensional
hypersurface in curved spacetime, the body's world tube. In the regime 
of static tides described previously, this hypersurface can be
foliated by $t=\mbox{constant}$ slices of the background spacetime,
and each leaf represents a closed two-surface, the body's boundary at 
each moment of time. This two-surface possesses a well-defined intrinsic
geometry, and a well-defined Ricci scalar that can be expressed as in
Eq.~(\ref{deformed_sphere2}). This construction supplies
$\delta {\cal R}$ with a precise meaning in general relativity, and
an interpretation for the left-hand side of Eq.~(\ref{h_Ndef2}). 

A precise meaning can also be assigned to $\delta {\cal R}$ in the
case of a deformed black hole. The tidal deformation of the event
horizon of a nonrotating black hole was examined in great generality
by Vega, Poisson, and Massey \cite{vega-poisson-massey:11}. Making use 
of the horizon's null generators to establish a coordinate system on
the horizon, the horizon's intrinsic geometry is described in terms of
a degenerate metric $\gamma_{AB}$ that is explicitly
two-dimensional. The Ricci curvature associated with this metric can
be expressed as in Eq.~(\ref{deformed_sphere2}) with $R = 2M$, thereby
providing a precise definition for $\delta {\cal R}$ in the case of a
black hole.   

With ${\cal E}_L$ and $\delta {\cal R}$ thus defined in general
relativity (for both material bodies and black holes),
Eq.~(\ref{h_Ndef2}) provides $h_\ell$ with a proper
general-relativistic definition. Because the tidal moments and surface
curvature are both defined in a coordinate-invariant manner, this
property is necessarily shared by the surficial Love numbers. 

In the following we will endeavor to express $\delta {\cal R}$ in
terms of the perturbation of the four-dimensional spacetime metric, 
and to find a relation between $h_\ell$ and $k^{\rm el}_\ell$ that
generalizes Eq.~(\ref{hvsk_N}) to general relativity.   

\section{Perturbed hydrostatic equilibrium} 
\label{sec:hydrostatic} 

We begin with a computation of $\delta {\cal R}$ in the case of a
material body; the case of a black hole will be considered next. The
first step is to locate the deformed surface of the body, which
requires an understanding of the hydrostatic equilibrium of the
perturbed configuration.   

The background spacetime of the unperturbed body has a static and 
spherically-symmetric metric given by  
\begin{equation} 
ds^2 = -e^{2\psi}\, dt^2 + f^{-1}\, dr^2 + r^2\, d\Omega^2 
\label{back_g} 
\end{equation} 
with $f := 1-2m/r$, in which $\psi$ and $m$ depend on the 
radial coordinate $r$. The metric is a solution to the Einstein field
equations with a perfect-fluid energy-momentum tensor 
\begin{equation} 
T^{\alpha\beta} = (\mu+p) u^\alpha u^\beta + p g^{\alpha\beta}, 
\label{em_tensor} 
\end{equation} 
in which $u^\alpha$ is the fluid's velocity field, $p$ is the
pressure, and $\mu = \rho + \epsilon$ is the total energy density,
decomposed into a rest-mass density $\rho$ and a density of internal
(thermodynamic) energy $\epsilon$. For a static configuration the only
nonvanishing component of the velocity vector is $u^t = e^{-\psi}$,
and the configuration is determined by the field equations 
\begin{equation} 
m' = 4\pi r^2 \mu, \qquad 
\psi' =\frac{m + 4\pi r^3 p}{r^2 f}, 
\label{back_EFE} 
\end{equation} 
in which a prime indicates differentiation with respect to $r$. In the
vacuum exterior the field equations produce the Schwarzschild solution
$m = M = \mbox{constant}$ and $e^{2\psi} = 1-2M/r$. The body's surface
is situated at $r=R$, as determined by the condition $p(R) = 0$. 

The equation of hydrostatic equilibrium, $(\mu + p) a_\alpha
+ \partial_\alpha p = 0$, in which $a_\alpha := u^\beta \nabla_\beta
u_\alpha$ is the fluid's covariant acceleration (with $a_r = \psi'$ as
its only nonvanishing component) reduces to 
\begin{equation} 
p' = -(\mu+p) \psi' = -\frac{(\mu+p)(m + 4\pi r^3 p)}{r^2 f}; 
\label{TOV}
\end{equation} 
this is the famous TOV (Tolman-Oppenheimer-Volkov) equation.  

The perturbed spacetime has a metric $g_{\alpha\beta} +
p_{\alpha\beta}$, and the perturbed fluid has an energy density 
$\mu + \delta \mu$, a pressure $p + \delta p$, a velocity $u^\alpha +
\delta u^\alpha$, and an acceleration $a_\alpha + \delta a_\alpha$. The
perturbed configuration satisfies the equation of hydrostatic
equilibrium  
\begin{equation} 
(\mu+p) \delta a_\alpha + (\delta \mu + \delta p) a_\alpha 
+ \partial_\alpha \delta p = 0, 
\label{hydro} 
\end{equation} 
and for a static configuration we find that $\delta a_t = 0$, 
$\delta a_r = -\frac{1}{2} \partial_r (e^{-2\psi} p_{tt})$, and 
$\delta a_A = -\frac{1}{2} \partial_A (e^{-2\psi} p_{tt})$. The radial
component of Eq.~(\ref{hydro}) is 
\begin{equation} 
-\frac{1}{2} (\mu+p) \partial_r \bigl( e^{-2\psi} p_{tt} \bigr) 
+ \psi' (\delta\mu + \delta p) + \partial_r \delta p = 0, 
\label{hydro_radial} 
\end{equation} 
while the angular components are 
\begin{equation} 
\partial_A \biggl[ -\frac{1}{2} (\mu+p) e^{-2\psi} p_{tt} 
+ \delta p \biggr] = 0. 
\end{equation} 
Integrating the second equation and inserting the result within the
first allows us to express $\delta \mu$ and $\delta p$ in terms of the
metric perturbation $p_{tt}$. We obtain 
\begin{equation} 
\delta \mu = -r \mu' F, \qquad 
\delta p = -rp' F, 
\label{fluid_pert} 
\end{equation} 
in which $F$ is defined by 
\begin{equation} 
r \psi' F := \frac{1}{2} e^{-2\psi} p_{tt}. 
\label{F_def} 
\end{equation} 

The solutions (\ref{fluid_pert}) to the equation of hydrostatic
equilibrium provide $F$ with a clear physical meaning. Suppose
that $r = r_0$ describes a surface of constant density $\mu = \mu_0$
in the spherical, unperturbed configuration. Suppose also that 
$r = r_0 + \delta r$ describes the deformed surface 
$\mu + \delta \mu = \mu_0$ in the perturbed configuration, and that we
wish to find the displacement $\delta r$. We have 
$\mu_0 = \mu(r_0 + \delta r) + \delta\mu(r_0) 
= \mu_0 + \mu'(r_0) \delta r + \delta\mu(r_0)$, and
inserting Eq.~(\ref{fluid_pert}) produces $\delta r = r_0 F(r_0)$. 
This calculation reveals that a spherical surface
$\mu = \mbox{constant}$ at radius $r$ is deformed to a surface 
$r + \delta r$ by the perturbation, with $\delta r = r F$. The same 
statement applies to a surface $p = \mbox{constant}$, and it applies 
in particular to $p = 0$, which marks the boundary of the fluid
configuration --- the body's surface. This implies that the
deformation of the body's surface is described by   
\begin{equation} 
\frac{\delta R}{R} = F(R,\theta^A), 
\label{surface_deformation} 
\end{equation} 
with $F$ defined by Eq.~(\ref{F_def}). Taking into account that
$e^{2\psi} = f$ and $\psi' = M/(R^2 f)$ on the boundary, we arrive at    
\begin{equation} 
F(R,\theta^A) = \frac{R}{2M} p_{tt}(R,\theta^A),  
\label{F_boundary} 
\end{equation} 
an explicit expression for $\delta R/R$ in terms of the metric
perturbation. 

The computation of $\delta {\cal R}$ can now be completed. It is easy
to show that to first order in the perturbation, the induced metric on
a two-surface  $t = \mbox{constant}$, $r = R(1 + F)$ in a spacetime
with metric $g_{\alpha\beta} + p_{\alpha\beta}$ is given by
Eq.~(\ref{deformed_sphere1}) with 
\begin{equation} 
\delta \Omega_{AB} = 2 F \Omega_{AB} + R^{-2} p_{AB},
\label{deformed_sphere4}
\end{equation} 
in which $F$ and $p_{AB}$ are evaluated at $r = R$ and expressed as
functions of the angular coordinates $\theta^A$. Inserting this within
Eq.~(\ref{deformed_sphere3}) gives 
\begin{equation} 
\delta {\cal R} = -2(D^2+2) F + \frac{1}{R^2} \bigl[ D^A D^B  
- (D^2 + 1) \Omega^{AB} \bigr] p_{AB}, 
\label{dR1} 
\end{equation} 
where we adopt the same conventions regarding the raising of angular
indices as described below Eq.~(\ref{deformed_sphere3}). 

At this stage it is helpful to introduce a decomposition of
$p_{\alpha\beta}$ into tensorial spherical harmonics. The relevant
fields are (we adopt the notation of Ref.~\cite{martel-poisson:05}) 
\begin{subequations} 
\label{sphharm_decomp} 
\begin{align} 
p_{tt} &= \sum_{\ell m} h^{\ell m}_{tt} Y^{\ell m}, \\ 
p_{AB} &= r^2 \sum_{\ell m} \bigl( K^{\ell m} \Omega_{AB} Y^{\ell m} 
+ G^{\ell m} Y_{AB}^{\ell m} \bigr) 
+ \sum_{\ell m} h_2^{\ell m} X_{AB}^{\ell m}, 
\end{align} 
\end{subequations} 
in which $h^{\ell m}_{tt}$, $K^{\ell m}$, $G^{\ell m}$, $h^{\ell m}_2$
depend on $r$ only, and 
\begin{subequations} 
\label{tensor_harmonics} 
\begin{align} 
Y_{AB}^{\ell m} &= \biggl[ D_A D_B + \frac{1}{2} \ell(\ell+1)
\Omega_{AB} \biggr] Y^{\ell m}, \\  
X_{AB}^{\ell m} &= -\frac{1}{2} \bigl( \varepsilon_A^{\ C} D_B 
+ \varepsilon_B^{\ C} D_A \bigr) D_C Y^{\ell m} 
\end{align} 
\end{subequations} 
are (tracefree) tensorial harmonics of even and odd parity,
respectively; $\varepsilon_{AB}$ is the Levi-Civita tensor on the unit
two-sphere, with components $\varepsilon_{\theta\phi} =
-\varepsilon_{\phi\theta} = \sin\theta$. Making the substitutions in
Eq.~(\ref{dR1}), and taking into account Eq.~(\ref{F_boundary}), we
arrive at  
\begin{equation} 
\delta {\cal R} = \sum_{\ell m} (\ell-1)(\ell+2) \biggl[ 
\frac{R}{M} h^{\ell m}_{tt} + K^{\ell m} 
+ \frac{1}{2} \ell(\ell+1) G^{\ell m} \biggr] Y^{\ell m}  
\label{dR2} 
\end{equation} 
after evaluating all perturbation fields at $r=R$, and 
after simplification of the angular operations on the spherical
harmonics. These manipulations make use of $D^2 Y^{\ell m} =
-\ell(\ell+1) Y^{\ell m}$, $D^A D^B Y^{\ell m}_{AB} = \frac{1}{2}
(\ell-1) \ell (\ell+1) (\ell+2) Y^{\ell m}$, and $D^A D^B X_{AB} =
0$. Notice that $\delta {\cal R}$ involves only the even-parity
fields $h^{\ell m}_{tt}$, $K^{\ell m}$, and $G^{\ell m}$; there is no
contribution from the odd-parity field  $h^{\ell m}_2$. This was to be
expected, because $\delta {\cal R}$ is a scalar that can only be
constructed from even-parity perturbations.  

We may now prove that $\delta {\cal R}$ is a gauge-invariant
quantity. Under a small coordinate transformation 
$x^\alpha_{\rm new} = x^\alpha_{\rm old} + \Xi^\alpha$ the metric 
perturbation transforms as $p^{\rm new}_{\alpha\beta} = 
p^{\rm old}_{\alpha\beta} - \nabla_\alpha \Xi_\beta 
- \nabla_\beta \Xi_\alpha$. Adopting the decompositions $\Xi_r =
\sum_{\ell m} \xi^{\ell m}_r Y^{\ell m}$ and 
$\Xi_A = \sum_{\ell m} \xi^{\ell m} D_A Y^{\ell m}$, we find that the
perturbation fields change according to 
\begin{subequations} 
\label{gauge_transf} 
\begin{align} 
h^{\rm new}_{tt} &= h^{\rm old}_{tt} + 2 e^{2\psi} \psi' \xi^r, \\ 
K^{\rm new} &= K^{\rm old} - \frac{2}{r} \xi^r 
+ \frac{\ell(\ell+1)}{r^2} \xi, \\ 
G^{\rm new} &= G^{\rm old} - \frac{2}{r^2} \xi. 
\end{align} 
\end{subequations} 
Evaluating this at $r=R$ and making the substitutions in
Eq.~(\ref{dR2}) reveals that $\delta {\cal R}^{\rm new} 
= \delta {\cal R}^{\rm old}$; the curvature perturbation is indeed
gauge invariant. This conclusion was expected, given that the surface
curvature is a meaningful, coordinate-independent, geometrical
quantity.  

\section{Curvature perturbation of a black hole} 
\label{sec:blackhole}

Equation (\ref{dR2}) describes the curvature perturbation of the
deformed surface of a material body. This will be related to the tidal
moments ${\cal E}_L$ in Sec.~\ref{sec:h_vs_k}, but before we proceed
we calculate $\delta {\cal R}$ for the event horizon of a deformed
black hole.   

Most of the work was carried out by Vega, Poisson, and Massey
\cite{vega-poisson-massey:11}, who examined the more general 
case of dynamical tides; here we specialize their results to the
regime of static tides. A general expression for $\delta {\cal R}$ is
given in their Eq.~(3.45), in terms of quantities defined in
Eqs.~(3.37) and (3.38); they have 
\begin{equation} 
\delta {\cal R} = \sum_{\ell m} (\ell-1)(\ell+2) \biggl[ 2 b^{\ell m} 
+ K^{\ell m} + \frac{1}{2} \ell(\ell+1) G^{\ell m} \biggr] Y^{\ell m},  
\label{dR3} 
\end{equation}  
with $K^{\ell m}$ and $G^{\ell m}$ evaluated at $r=2M$ as functions of
$v$, and 
\begin{equation} 
b^{\ell m}(v) := \kappa \int_v^\infty e^{-\kappa(v'-v)} 
h^{\ell m}_{tt}(v',2M)\, dv';  
\label{b_def} 
\end{equation} 
here $v$ is the Eddington-Finkelstein advanced time defined by 
$dv := dt + f^{-1} dr$, which is well behaved on the event horizon,
and $\kappa := 1/(4M)$ is the horizon's surface gravity. In the regime
of  static tides the time scale associated with variations in 
$h^{\ell m}_{tt}$ is very long compared with $\kappa^{-1}$, and the 
integral of Eq.~(\ref{b_def}) can be approximated by 
$b^{\ell m} = h^{\ell m}_{tt}$. Making the substitution in
Eq.~(\ref{dR3}) returns  
\begin{equation} 
\delta {\cal R} = \sum_{\ell m} (\ell-1)(\ell+2) \biggl[ 
2 h^{\ell m}_{tt}  + K^{\ell m} 
+ \frac{1}{2} \ell(\ell+1) G^{\ell m} \biggr] Y^{\ell m},  
\label{dR4} 
\end{equation}  
in which all perturbation fields are evaluated at $r=2M$.

Equation (\ref{dR4}) can be compared with Eq.~(\ref{dR2}), and this
reveals that the black-hole result can be obtained as a special case
of the material-body result by setting $R=2M$. This is a remarkable
outcome, given that the surface of a material body and the event
horizon of a black hole are physically extremely
different. Nevertheless, we have established that the surface
deformation of material bodies and black holes can be described
meaningfully in a unified manner, in terms of their surface
curvature. 

\section{Relation between surficial and gravitational Love numbers} 
\label{sec:h_vs_k} 

Because the perturbation fields $h^{\ell m}_{tt}$, $K^{\ell m}$, and
$G^{\ell m}$ are continuous at $r=R$, the curvature perturbation
of Eq.~(\ref{dR2}) can be evaluated by relying on the external
solutions of the perturbation equations instead of the internal
solutions. And because $\delta {\cal R}$ is gauge invariant, the
computation can be carried out in any convenient gauge; we adopt the
simplest choice of the Regge-Wheeler gauge. The solutions to
the vacuum external problem are well known \cite{zerilli:70}, and can
be directly imported here.  

Specifically we adapt the discussion contained in Sec.~III of
Binnington and Poisson \cite{binnington-poisson:09}, which was framed 
in the light-cone gauge, to the Regge-Wheeler gauge. We find that the
external fields are given by   
\begin{subequations}
\label{external_slns1}
\begin{align} 
h^{\ell m}_{tt} &= -\frac{2}{(\ell-1)\ell} r^\ell f^2   
\bigl[ A_1 + 2k^{\rm el}_\ell (R/r)^{2\ell+1} B_1 \bigr] 
{\cal E}^{(\ell)}_m, \\ 
K^{\ell m} &= -\frac{2}{(\ell-1)\ell}  r^\ell  
\bigl[ A_2 + 2k^{\rm el}_\ell (R/r)^{2\ell+1} B_2 \bigr] 
{\cal E}^{(\ell)}_m, \\ 
G^{\ell m} &= 0, 
\end{align}
\end{subequations}  
where $f = 1-2M/r$, $k^{\rm el}_{\ell}$ are the electric-type
gravitational Love numbers, and  
\begin{subequations} 
\label{external_slns2} 
\begin{align} 
A_1 &=F(-\ell,-\ell+2;-2\ell;2M/r), \\ 
B_1 &= F(\ell+1,\ell+3;2\ell+2;2M/r), \\ 
A_2 &= \frac{\ell+1}{\ell-1} F(-\ell,-\ell;-2\ell;2M/r) 
- \frac{2}{\ell-1} F(-\ell,-\ell-1;-2\ell;2M/r), \\ 
B_2 &= \frac{\ell}{\ell+2} F(\ell+1,\ell+1;2\ell+2;2M/r) 
+ \frac{2}{\ell+2} F(\ell+1,\ell;2\ell+2;2M/r) 
\end{align} 
\end{subequations} 
are functions of $2M/r$ expressed in terms of hypergeometric
functions.  

Making the substitutions in Eq.~(\ref{dR2}) and incorporating
Eq.~(\ref{STF_Ylm}) returns  
\begin{equation} 
\delta {\cal R} = -2 \sum_\ell \frac{\ell+2}{\ell} 
\biggl[ f^2 \bigl( A_1 + 2k^{\rm el}_\ell B_1 \bigr)
+ \frac{M}{R} \bigl( A_2 + 2k^{\rm el}_\ell B_2 \bigr) \biggr] 
\frac{R^{\ell+1}}{M} {\cal E}_L \Omega^L. 
\end{equation} 
Comparing with Eq.~(\ref{h_Ndef2}) reveals that the quantity within
square brackets can be identified with $h_\ell$. After simplification
we arrive at  
\begin{equation} 
h_\ell = \Gamma_1 + 2  \Gamma_2\, k^{\rm el}_{\ell}
\label{hvsk_rel} 
\end{equation} 
with 
\begin{subequations} 
\label{Gamma} 
\begin{align} 
\Gamma_1 &= \frac{\ell+1}{\ell-1} (1-M/R) F(-\ell,-\ell;-2\ell;2M/R) 
- \frac{2}{\ell-1} F(-\ell,-\ell-1;-2\ell;2M/R), \\ 
\Gamma_2 &= \frac{\ell}{\ell+2} (1-M/R) F(\ell+1,\ell+1;2\ell+2;2M/R) 
+ \frac{2}{\ell+2} F(\ell+1,\ell;2\ell+2;2M/R). 
\end{align} 
\end{subequations} 
To obtain these expressions we made use of standard relations
among contiguous hypergeometric functions. The properties of 
$\Gamma_1$ and $\Gamma_2$ were described in Sec.~\ref{sec:intro}.  

Equation (\ref{hvsk_rel}) provides a complete solution to the problem
of determining the surficial Love numbers of material bodies and black
holes in general relativity. A computation of $h_\ell$ is exceedingly
simple once $k^{\rm el}_\ell$ and $M/R$ have been obtained for a
selected body.  

\section{Surficial Love numbers of black holes} 
\label{sec:black} 

To compute $h_\ell$ for black holes we rely on the fact that  
$k^{\rm el}_\ell = 0$ and evaluate Eq.~(\ref{hvsk_rel})
explicitly. This requires the evaluation of hypergeometric
functions  at $2M/R = 1$, which is achieved with the Chu-Vandermonde
identity [Eq.~(15.4.24) of Ref.~\cite{NIST:10}], 
$F(-\ell,b;c;1) = (c-b)_\ell/(c)_\ell$, where 
$(a)_\ell = a (a+1) \cdots (a+\ell-1)$ is the Pochhammer symbol. With  
simple manipulations we find that 
$F(-\ell,-\ell;-2\ell;1) = \ell!^2/(2\ell)!$,  
$F(-\ell,-\ell-1;-2\ell;1) = 0$, and substitution within 
Eq.~(\ref{hvsk_rel}) yields 
\begin{equation} 
h_\ell = \frac{\ell+1}{2(\ell-1)} \frac{\ell!^2}{(2\ell)!}. 
\label{h_BH} 
\end{equation} 
Equation (\ref{h_BH}) agrees with the result of Damour and Lecian
\cite{damour-lecian:09}. 

\section{Computation of gravitational Love numbers} 
\label{sec:recipe} 

In this section we summarize and simplify the recipe concocted by
Damour and Nagar \cite{damour-nagar:09} to calculate the gravitational 
Love numbers of a perfect-fluid body in general relativity. Since most
of this material can be extracted from Damour and Nagar and
Binnington and Poisson \cite{binnington-poisson:09}, we provide very
few derivations.  

The electric-type Love numbers $k^{\rm el}_\ell$ are determined by
integrating the equations governing the even-parity perturbations of a
spherical body. After incorporating the solutions to the equation of
hydrostatic equilibrium obtained in Sec.~\ref{sec:hydrostatic},
performing a decomposition in spherical harmonics, and adopting the
Regge-Wheeler gauge, the field equations for the metric perturbation
inside the body give rise to a decoupled differential equation for
$h^{\ell m}_{tt}$,  
\begin{equation} 
r^2 h''_{tt} + A r h'_{tt}  - B h_{tt} = 0, 
\label{pert_even1} 
\end{equation} 
with 
\begin{subequations} 
\label{pert_even2} 
\begin{align} 
A &= \frac{2}{f} \Bigl[ 1 - 3m/r - 2\pi r^2 (\mu + 3p) \Bigr], \\ 
B &= \frac{1}{f} \Bigl[ \ell(\ell+1) 
- 4\pi r^2 (\mu + p) (3 + d\mu/dp) \Bigr].
\end{align} 
\end{subequations} 
This equation is integrated from $r = 0$, near which the solution
behaves as $h_{tt} \propto r^\ell$. The internal solution is matched
to the external solution (\ref{external_slns1}) at $r=R$, and the
matching returns $k^{\rm el}_\ell$. 

An efficient way to implement this procedure is to introduce the
logarithmic derivative $\eta := r h'_{tt}/h_{tt}$ and to recast
Eq.~(\ref{pert_even1}) as 
\begin{equation} 
r \eta' + \eta(\eta-1) + A \eta - B = 0. 
\label{pert_even3} 
\end{equation} 
The main advantage of this formulation is that it is better
conditioned for numerical integration. It also eliminates the need to
determine the overall constant that multiplies $h_{tt}$ in the
original formulation. This equation also is integrated from $r = 0$, 
at which $\eta = \ell$, and the main outcome of the computation
is $\eta_{\rm s} := \eta(r=R)$, the surface value of the logarithmic
derivative. 

Calculating $rh'_{tt}/h_{tt}$ from Eq.~(\ref{external_slns1}) reveals
that the matching condition at $r=R$ is 
\begin{equation} 
\eta_{\rm s} = \ell + \frac{4M}{R-2M} 
+ \frac{ RA'_1 + 2k^{\rm el} \bigl[R B'_1 - (2\ell+1) B_1 \bigr] }
{ A_1 + 2 k^{\rm el} B_1 }, 
\end{equation} 
in which $A_1$ and $B_1$ are the hypergeometric functions introduced
in Eq.~(\ref{external_slns2}). Solving for $k^{\rm el}$ gives 
\begin{equation} 
2 k_\ell^{\rm el} = \frac{ R A'_1 - \bigl[ \eta_{\rm s} - \ell 
  - 4M/(R-2M) \bigr] A_1 }{ \bigl[ \eta_{\rm s} + \ell + 1 
  - 4M/(R-2M) \bigr] B_1 - R B'_1 }. 
\label{kel_sln} 
\end{equation} 
With $\eta_{\rm s}$ obtained from the integration of
Eq.~(\ref{pert_even3}), the electric-type Love number
$k^{\rm el}_\ell$ follows after evaluating a couple of hypergeometric
functions and their derivatives at $r=R$.   

A similar recipe can formulated for the magnetic-type Love numbers
$k^{\rm mag}_\ell$. Here the decoupled equation for the internal
perturbation involves the coefficients $h^{\ell m}_t$ of the
decomposition of $p_{tA}$ in odd-parity vectorial harmonics. It   
reads 
\begin{equation} 
r^2 h''_{t} - P r h'_{t}  - Q h_{t} = 0, 
\label{pert_odd1} 
\end{equation} 
with 
\begin{subequations} 
\label{pert_odd2} 
\begin{align} 
P &= \frac{4\pi r^2}{f} (\mu + p), \\ 
Q &= \frac{1}{f} \Bigl[ \ell(\ell+1) 
- 4m/r + 8\pi r^2 (\mu + p) \Bigr].
\end{align} 
\end{subequations} 
This equation is integrated from $r = 0$, near which the solution
behaves as $h_{t} \propto r^{\ell+1}$. The internal solution is
matched across $r=R$ to the external solution 
\begin{equation} 
h^{\ell m}_{t} = \frac{2}{3(\ell-1)\ell} r^{\ell+1}     
\biggl[ A_3 - 2 \frac{\ell+1}{\ell} k^{\rm mag}_\ell (R/r)^{2\ell+1}
B_3 \biggr] {\cal B}^{(\ell)}_m, 
\end{equation}  
where $k^{\rm mag}_\ell$ are the magnetic-type gravitational Love 
numbers,   
\begin{subequations} 
\begin{align} 
A_3 &=F(-\ell+1,-\ell-2;-2\ell;2M/r), \\ 
B_3 &= F(\ell-1,\ell+2;2\ell+2;2M/r), 
\end{align} 
\end{subequations} 
and ${\cal B}^{(\ell)}_m$ is the spherical-harmonic packaging of the
tidal moments ${\cal B}_L$, defined as in Eq.~(\ref{STF_Ylm}). 

The equivalent formulation in terms of $\kappa := rh'_t/h_t$ is 
\begin{equation} 
r \kappa' + \kappa(\kappa-1) - P \kappa - Q = 0, 
\label{pert_odd3} 
\end{equation} 
and the equation is integrated from $r=0$ with $\kappa = \ell + 1$ 
to $r = R$ at which $\kappa = \kappa_{\rm s}$. In this case the
matching condition at $r=R$ produces  
\begin{equation} 
2 \frac{\ell+1}{\ell} k_\ell^{\rm mag} = 
\frac{ RA'_3 - (\kappa_{\rm s} - \ell - 1) A_3 }
{ RB'_3 - (\kappa_{\rm s} + \ell) B_3 }. 
\label{kmag_sln} 
\end{equation} 

\begin{acknowledgments} 
We are grateful for discussions with Kent Yagi. This work was
supported by the Natural Sciences and Engineering Research Council of
Canada.    
\end{acknowledgments}    

\bibliography{../bib/master} 
\end{document}